\documentclass[runningheads,a4paper]{llncs}

\usepackage{amssymb}
\usepackage{graphicx}

\usepackage{url}
\urldef{\mailsa}\path|hkwak@qf.org.qa|
\urldef{\mailsb}\path|jeremyb@tid.es|
\newcommand{\keywords}[1]{\par\addvspace\baselineskip
\noindent\keywordname\enspace\ignorespaces#1}

\usepackage[usenames,dvipsnames]{xcolor}

\usepackage{soul}

\newcounter{NumberOfComments}
\stepcounter{NumberOfComments}

\newcounter{JBNumberOfComments}
\stepcounter{JBNumberOfComments}

\newcommand{\eat}[1]{}

\begin{document}

\mainmatter  

\title{Linguistic Analysis of Toxic Behavior in an Online Video Game}

\titlerunning{Linguistic Analysis of Toxic Behavior in an Online Video Game}

%
%
\author{Haewoon Kwak$^{\dag}$ \and Jeremy Blackburn$^{\ddag}$}
\authorrunning{Haewoon Kwak and Jeremy Blackburn}

\institute{$^{\dag}$Qatar Computing Research Institute, Doha, Qatar\\
\mailsa\\
$^{\ddag}$Telefonica Research, Barcelona, Spain\\
\mailsb\\
}

%
%

\maketitle

\begin{abstract}
In this paper we explore the \emph{linguistic} components of toxic behavior by using crowdsourced data from over 590 thousand cases of accused toxic players in  a popular match-based competition game, League of Legends.  We perform a series of linguistic analyses to gain a deeper understanding of the role communication plays in the expression of toxic behavior.  We characterize linguistic behavior of toxic players and compare it with that of typical players in an online competition game.  We also find empirical support describing how a player \emph{transitions} from typical to toxic behavior.  
Our findings can be helpful to automatically detect and warn players who may become toxic and thus insulate potential victims from toxic playing in advance.


\end{abstract}


\section{Introduction}\label{sec:intro}

Multiplayer games provide players with the thrill of \emph{true} competition.
Players prove themselves superior to other humans that exhibit \emph{dynamic} behavior far beyond that of any computer controlled opponent.
Additionally, some multiplayer games provide another wrinkle: teamwork.
Now, not only is it a test of skill between two individuals, but cooperation, strategy, and communication between teammates can ensure victory.
Unfortunately, the presence of teammates and their influence on victory and defeat can result in \emph{toxic} behavior.

Toxic behavior, also known as cyberbullying~\cite{barlinska2013cyberbullying}, griefing~\cite{Chesney09}, or online disinhibition~\cite{suler2004online}, is bad behavior that violates social norms, inflicts misery, continues to cause harm after it occurs, and affects an entire community.
The anonymity afforded by, and ubiquity of, computer-mediated-communication (CMC) naturally leads to hostility and aggressiveness~\cite{Chen09,Thompson96}.
A major obstacle in understanding toxic behavior is its subjective perception.
Unlike unethical behavior like cheating, toxic behavior is nebulously defined; toxic players themselves sometimes fail to recognize their behavior as toxic~\cite{Lin05}.
Nevertheless, because of the very real impact toxic behavior has on our daily lives, even outside of games, a deeper understanding is necessary.

To further our understanding, in this paper we explore the \emph{linguistic} components of toxic behavior.
Using crowdsourced data from over 590 thousand ``judicial trials'' of accused toxic players representing over 2.1 million matches of a popular match-based competition game, League of Legends\footnote{http://leagueoflegends.com}, we perform a series of linguistic analyses to gain a deeper understanding of the role communication plays in the expression of toxic behavior.  
In our previous work~\cite{blackburn2014stfu}, we found that offensive language is the most reported reason across all the three regions.  Also, in North America, verbal abuse is the second most reported reason.  In other words, linguistic components are a prime method of expressing toxicity.  

From our analyses we draw several findings.
First, the \emph{volume} of communication is not uniform throughout the length of the match, instead showing a bi-modal shape with peaks at the beginning and end of a match.  
By comparing the distribution of frequency of communications between normal players and toxic players, we find subtle differences.
Typical players chat relatively more at the beginning of a match, which is mainly for ice breaking, morale boosting, and sharing early strategic information.
In contrast, toxic players chat less at the beginning but constantly more than typical players after some time point, i.e. phase transition. 
Next, we find discriminative uni- and bi-grams used by typical and toxic players, as signatures of them, examine the differences, and show that certain bi-grams can be classified based on \emph{when} they appear in a match.  Temporal patterns of the linguistic signature of toxic players illustrate what kind of toxic playing happens as the match progresses.  
Deeper analysis of temporal analysis of words used by toxic and typical players reveals a more interesting picture.  We focus on how a player transitions to toxic by comparing the temporal usage of popular uni-grams between typical players and toxic players.

Our contribution is two-fold.  First, we characterize linguistic behavior of toxic players and compare it with that of typical players in online competition games.  Second, we find empirical support to describe how a player turns to be toxic.  
Our findings would be helpful to automatically detect and warn players who may turn to be toxic and thus save potential victims of toxic playing in advance.


\section{Dataset}\label{sec:dataset}

The League of Legends (LoL) is the most popular Multiplayer Online Battle Arena out today, and suffers from a high degree of toxic behavior.
The LoL Tribunal is a crowdsourced system for determining the guilt of players accused of tocix behavior.

We collected 590,311 Tribunal cases from the North America region representing a total of 2,107,522 individual matches.
Each Tribunal case represents a single player and includes up to 5 matches in which he was accused of toxic behavior.
In LoL players can communicate via chat, which is ostensibly used to share strategic plans and other important information during the game.
However, chat is also a prime vector for exhibiting toxic behavior.
Thus, although a variety of information is presented to Tribunal reviewers~\cite{blackburn2014stfu}, in this paper we focus exclusively on the in-game \emph{chat logs}.


We extract 24,039,184 messages from toxic players and 33,252,018 messages from typical players.
Because the teammates of toxic players are directly impacted by toxic playing and readily express aggressive reactions to a toxic player, we define \emph{typical players} as the set of players on the opposite team when none of them report the toxic player.

Before continuing, we report some basic statistics about the size of vocabulary and the length of messages.  We found 1,042,940 unique tokens in toxic player messages and 1,176,356 unique tokens in typical player messages.
While typical players send 38\% more messages than toxic players, the messages are composed of only 13\% more unique tokens.
Interestingly, we find that toxic players send longer messages than typical players; the average number of words per message is 3.139 and 2.732 for toxic and typical players, respectively.


\section{Chat Volume over a Match}\label{sec:volume}

We begin our analysis by exploring chat volume over time.
A LoL match can be broken up into logical stages.
First is the \emph{early game} (also known as the ``laning phase''), where characters are low level and weak.
In the early game, players expend great effort towards ``farming'' computer controlled minions to gain experience and gold, with aggressive plays against the other team usually coming as the result of an over extension or other mistake.
As players earn gold and experience, they level up and become stronger, and the match transitions to the \emph{mid game}.
During the mid game, players become more aggressive and tend to group up with teammates to make plays on their opponents.
Finally, once players are reaching their maximum power levels, the match transitions into the \emph{end game}, where teams will group together and make hard pushes towards taking objectives and winning the match.

\begin{figure} [t!]
  \begin{center}
  \includegraphics[scale=0.35]{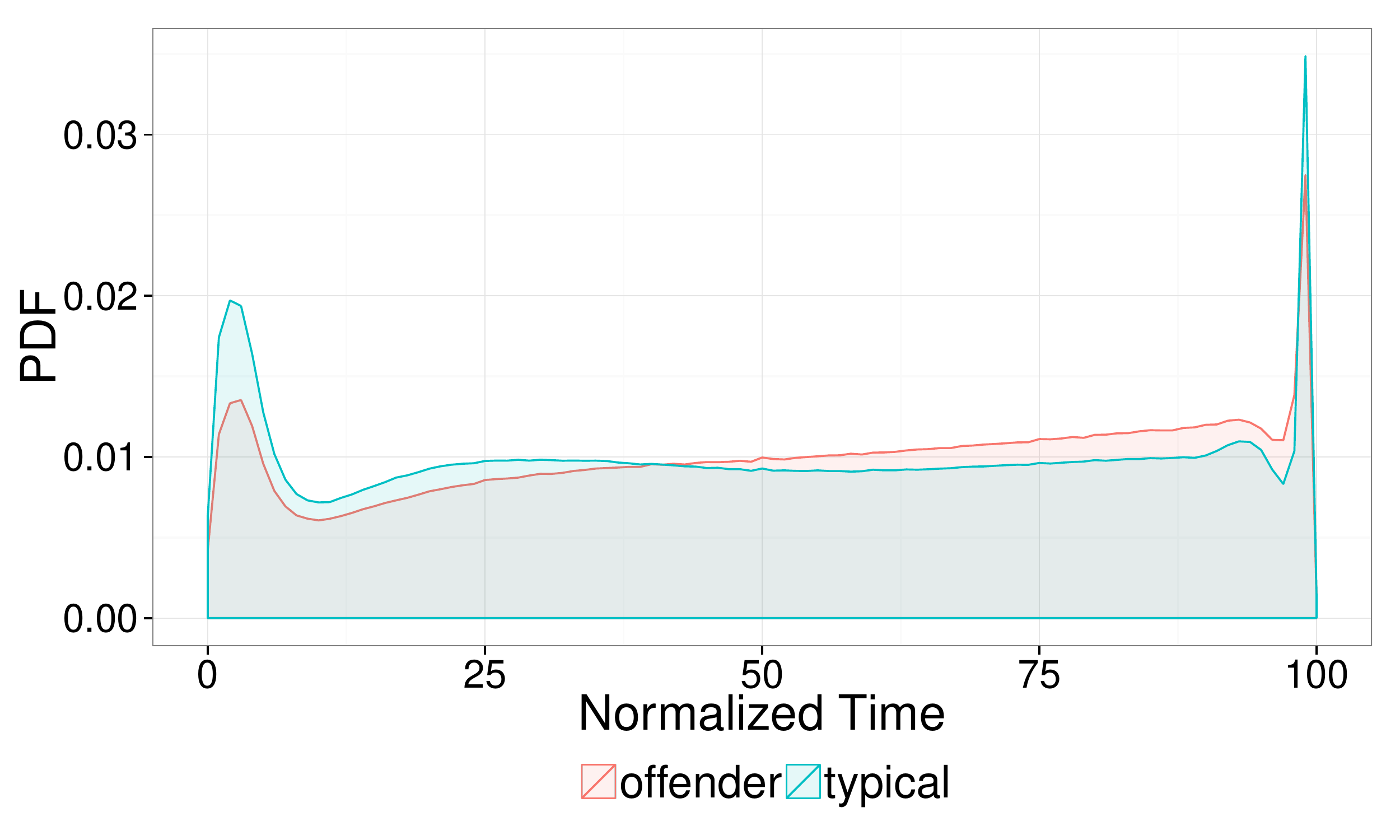}
  \caption{Change of chat volume during a match.}
  \vspace{-5mm} 
  \label{fig:whole_chat_timeline}
  \end{center}
\end{figure}

While these phases are not dictated by the programming of LoL, and thus there is no hard cut off point for when the transitions between phases occur, we suspect that each phase has an associated pattern of communication.
Thus, in Figure~\ref{fig:whole_chat_timeline} we plot the density of chat messages written by toxic and typical players as a function of the normalized time during a match.
The plot confirms our suspicions: communication is not uniform throughout the match.
Instead, we see three distinct levels of communication, likely corresponding to the three phases of a match, with relative peaks at the beginning and end of the match.

This finding can be explained with a deeper understanding of how a LoL match progresses.
As mentioned above, in the early game players are relatively weak and must focus on farming for resources.
Early game farming occurs via players choosing one of three lanes to spend their time in.
The lanes are quite far from each other on the map (10+ seconds or so to travel between them) and thus players on the same team tend to be relatively isolated from each other.
To take advantage of this isolation, and to get an early lead, players might roam from the lane they chose to play in to another lane.
In turn, this provides their teammate in the other lane with a numbers advantage over opposing player in the lane.
Colloquially, this roaming to provide a temporary numbers advantage is known as a ``gank.''
To deal with ganks in the early game, players tend to communicate via chat when the opposing player in their lane has gone missing.

As the match transitions to mid game, teammates start grouping up.
Since they are no longer so isolated the fear of ganks dissipates, and the need to communicate missing players diminishes.
Additionally, since teammates are grouped together, they are seeing the same portion of the map, and there is not really that much additional information they can convey to each other.

Finally, as late game comes around, teams must focus and work together to complete objectives and win the match.
In practice, this might involve coming to agreement on a final ``push'' for an objective, or agreeing on which lane the team should travel down.
Also, there are some customs in e-sports, saying `gg (good game)' at the end of the game.  The sharp spikes contain those messages as well.
While this might explain some of the spike seen at the end of Figure~\ref{fig:whole_chat_timeline} another, simpler explanation is that players are simply communicating their (dis)pleasure in winning or losing the match.

A more interesting finding is the subtle difference in the distributions of typical and toxic players.  At the early stage we see more active communication by normal players.  We suppose that it includes all the messages for ice breaking or cheering (e.g. gl (good luck) or hf (have fun)).  However, at some point after the short period, toxic players begin to chat more than typical players and keep such pattern until by the last stage.  At the last stage of the match, typical players again chat more socially, for example, sending smile emoticons, which are :D or :), and also saying gg, as we mentioned.  The transition point, where the distribution of toxic players cross over that of typical players, is a basis of our further analysis in Section~\ref{sec:long}.


\section{Discriminative Words of Toxic and Typical Players}

The linguistic approach to the chat log characterizes toxic players with context.  
We conduct $n$-gram analysis because it is intuitive and straightforward.  
We filter the stopwords and then count the frequency of uni- and bi-grams from the chat log involving toxic reports of either verbal abuse or offensive language. 

\begin{figure} [t!]
  \begin{center}
  \includegraphics[scale=0.2]{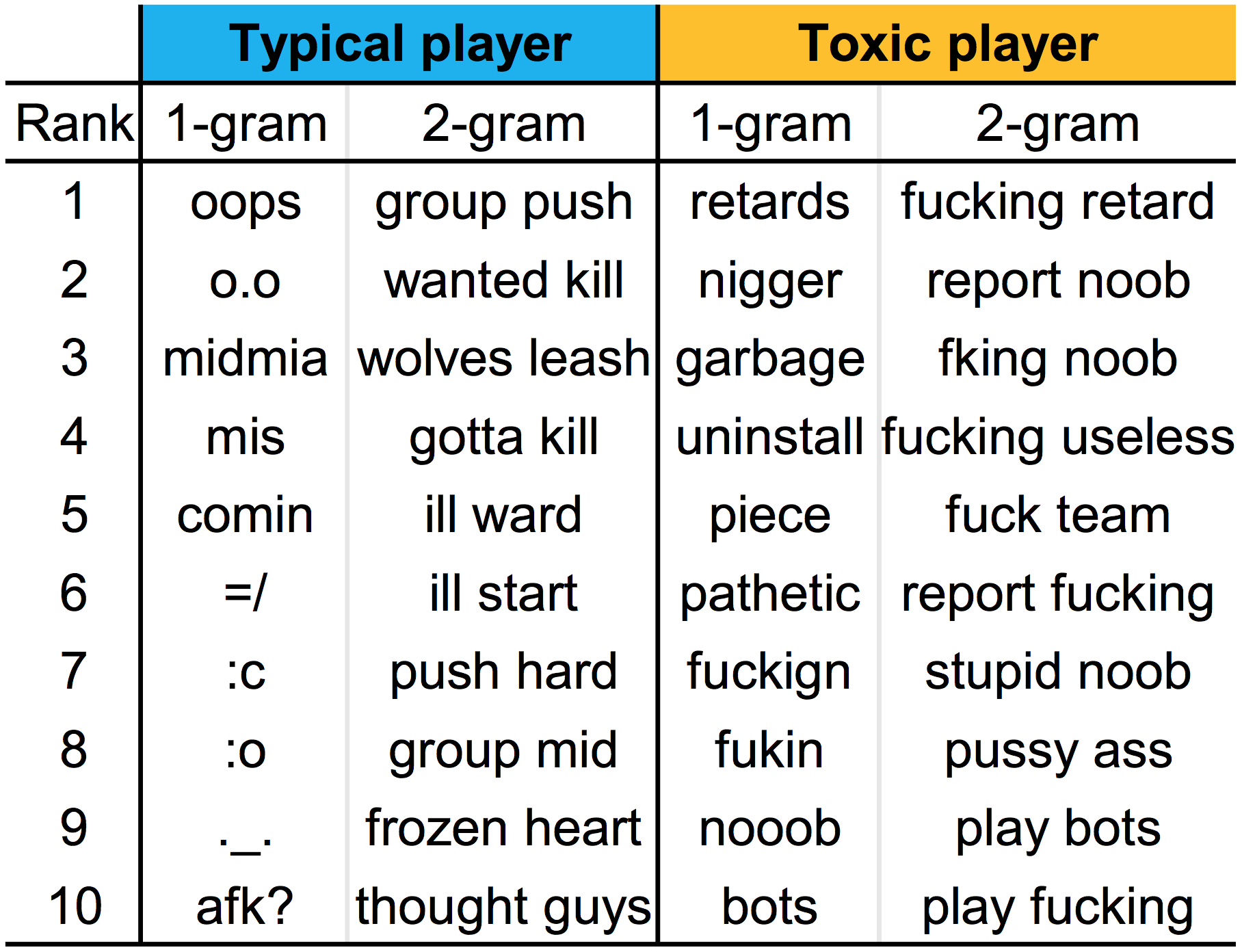}
  \caption{Top 10 discriminative uni- and bi-grams}
  \vspace{-5mm} 
  \label{fig:top10_bigram}
  \end{center}
\end{figure}

In order to find discriminative $n$-grams of toxic players we need a reference for comparison.  
We conduct the same $n$-gram analysis from enemy's chat log when verbal abuse or offensive language is $not$ reported from the enemies.  We consider it as a normal conversation among players and call those enemies typical players.
We create the top 1,000 uni- and bi-grams for toxic and typical players, respectively.  
We find 867 uni- and 748 bi-grams in common.  Then we obtain 133 non-overlapped uni- and 252 bi-grams for toxic and typical players; they appear only in either toxic or typical players.
We define them as discriminative uni- and bi-gram for toxic and typical players, respectively.  

Figure~\ref{fig:top10_bigram} shows top 10 discriminative uni- and bi-grams of toxic and typical players.
Top 10 discriminative uni- and bi-grams of toxic players are filled with bad words.  That is, Riot Games does not offer even the basic level of bad word filtering, and such bad words can be used as the signatures of toxic players who used verbal abuse or offensive language.
We find that several discriminative bi-grams of typical players are about strategies, while most of toxic players' bi-grams are bad words.  We note that some variations of `fucking' are discriminative uni-grams but `fucking' itself is not.  It means that `fucking' is often used not only by toxic players but also by typical players as well.  This shows the difficulties of filtering bad words by a simple dictionary-based approach. 

\begin{figure} [hbt!]
  \begin{center}
  \includegraphics[scale=0.3]{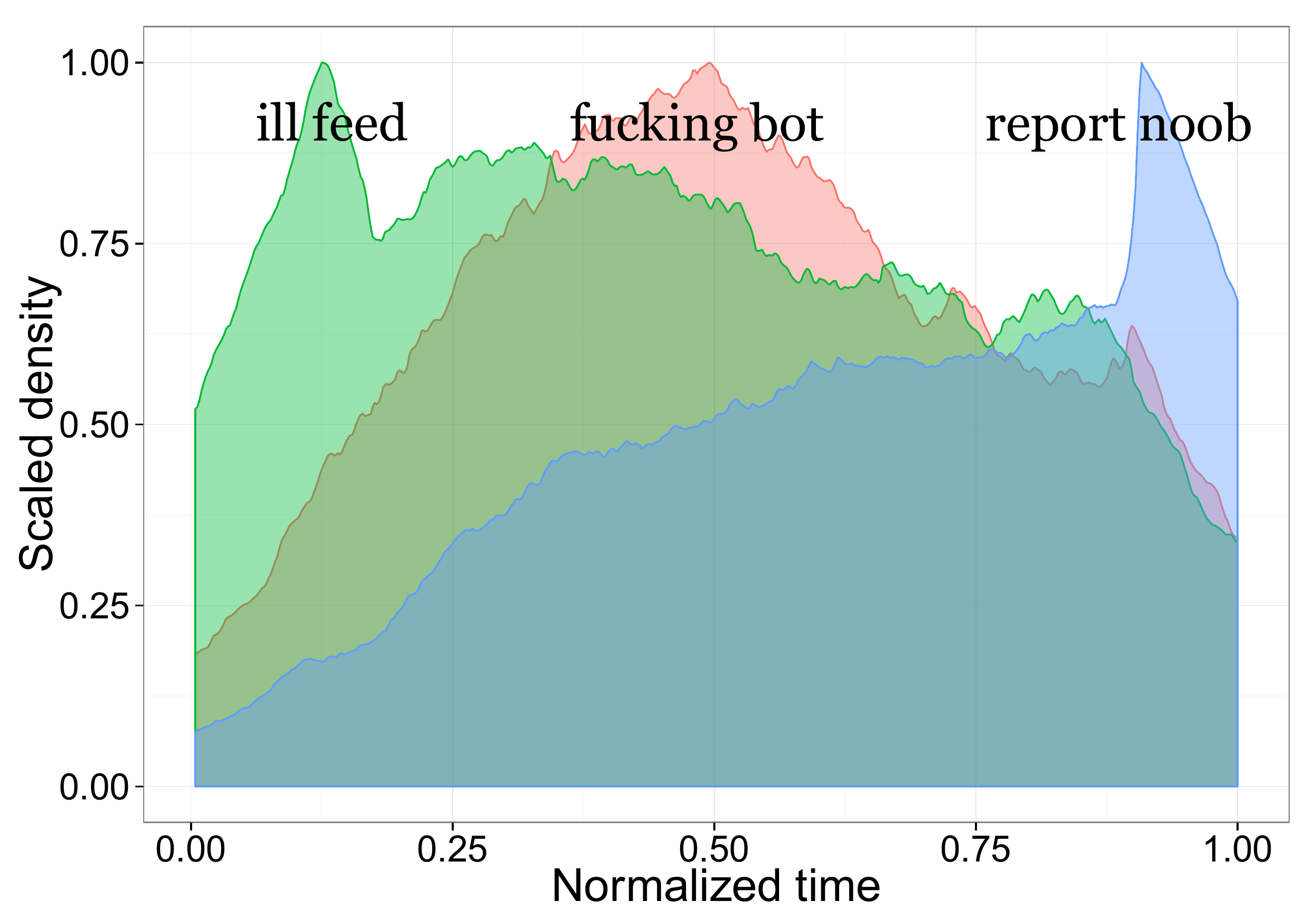}
  \caption{Example of early-, mid-, and late-bi-gram}
  \vspace{-5mm} 
  \label{fig:timeline_chat}
  \end{center}
\end{figure}

As the next step of the linguistic approach, we are interested in \emph{when verbal abuse occurs} from a temporal perspective during a match.  
We divide 252 discriminative bi-grams of toxic players into three classes, early-, mid-, and late-bi-grams, based on when their highest frequencies occur.

Figure~\ref{fig:timeline_chat} presents an example of three temporal classes of bi-grams.  Interestingly, 209 (82.94\%) out of 252 bi-grams are late-bi-gram.
The early-bi-gram ``ill feed'' is a domain specific example of toxic behavior.
In LoL, one of the ways players earn gold and experience during a match is by killing players on the opposite team.
Intentional feeding is when a player deliberately allows the other team to kill them, thus ``feeding'' the enemies with gold and experience, in turn allowing them to become quite powerful.

The mid-bi-gram ``fucking bot'' is the toxic player expressing his displeasure for the performance of the bottom lane.
The bottom lane is usually manned by characters that have a primarily late-game presence, and thus being behind during the mid-game has a significant impact on the remainder of the match.

Most verbal abuse of toxic players occurs in the late stage of the game.
For example, ``report noob'' is the toxic player requesting that the rest of his team report a player (the ``noob'') that he singled out for his ire.
We believe the most likely explanation for this is that verbal abuse is most likely a response to losing a game, which is often not apparent until the late-game.
For example, consider a scenario where one player on the team has a bad game, perhaps making poor decisions resulting in the enemy team becoming quite strong.
In the early-, and even mid-game phases, a toxic player might still be able to hold his own, however, when the enemy team groups up and makes coordinated pushes in the late-game, their relative strength will often result in quick and decisive victories in team-fights.
If toxic playing can be detected in real-time, we could protect potential victims from verbal violence, for example 
via alerts or simply not delivering such messages.

Temporal dynamics of bi-grams might help to create a mental model of toxic players.  For instance, 10 bi-grams containing `bot' are divided into 1 early-bi-gram, 5 mid-bi-grams, and 4 late-bi-grams.  Through manual inspection, we confirm that the early-bi-gram (`go bot') is strategic and non-aggressive, the mid-bi-grams are cursing, and the late-bi-grams are blaming the result of the match on the bot player(s).
This provides us with a rough idea of how toxic players might behave and think over time: initially they have a similar mindset as typical players, but, as the game plays out contrary to their desires, they grow increasingly aggressive, eventually lashing out with purely abusive language.
We leave more sophisticated modeling of toxic players' thought process as future work.


\section{Phase Transition of Toxic Players}\label{sec:long}

In the previous section we recognize which words are exclusively used by toxic and normal players.
However, some words are used by both toxic players and normal players.
For these, the emerging patterns in a temporal sense could be quite different.
If we assume that toxic players exhibit toxic behavior in reaction to certain events happening during the match, then the linguistic behavior of such toxic players \emph{should be the same as typical players before those events happen}.  

To validate the above hypothesis, we conduct the following experiment which is focused on finding some words that are not used after some time point by toxic players, while they are continuously used by normal players.
We extract the top 30 uni-gram at every normalized time unit (ranging from 0 to 100) for toxic players and normal players, respectively.  Since top 30 uni-grams are quite stable during the match, we obtain unique 80 uni-gram for toxic players and 91 uni-grams for normal players.  We first observe that toxic players have slightly smaller vocabularies than that by normal players.  For each of these uni-grams, we compute the normalized time of last use by toxic players and normal players, respectively. Finally, we compute the difference of the last used time between toxic and normal players for the common uni-grams.

\begin{figure} [t!]
  \begin{center}
  \includegraphics[scale=0.2]{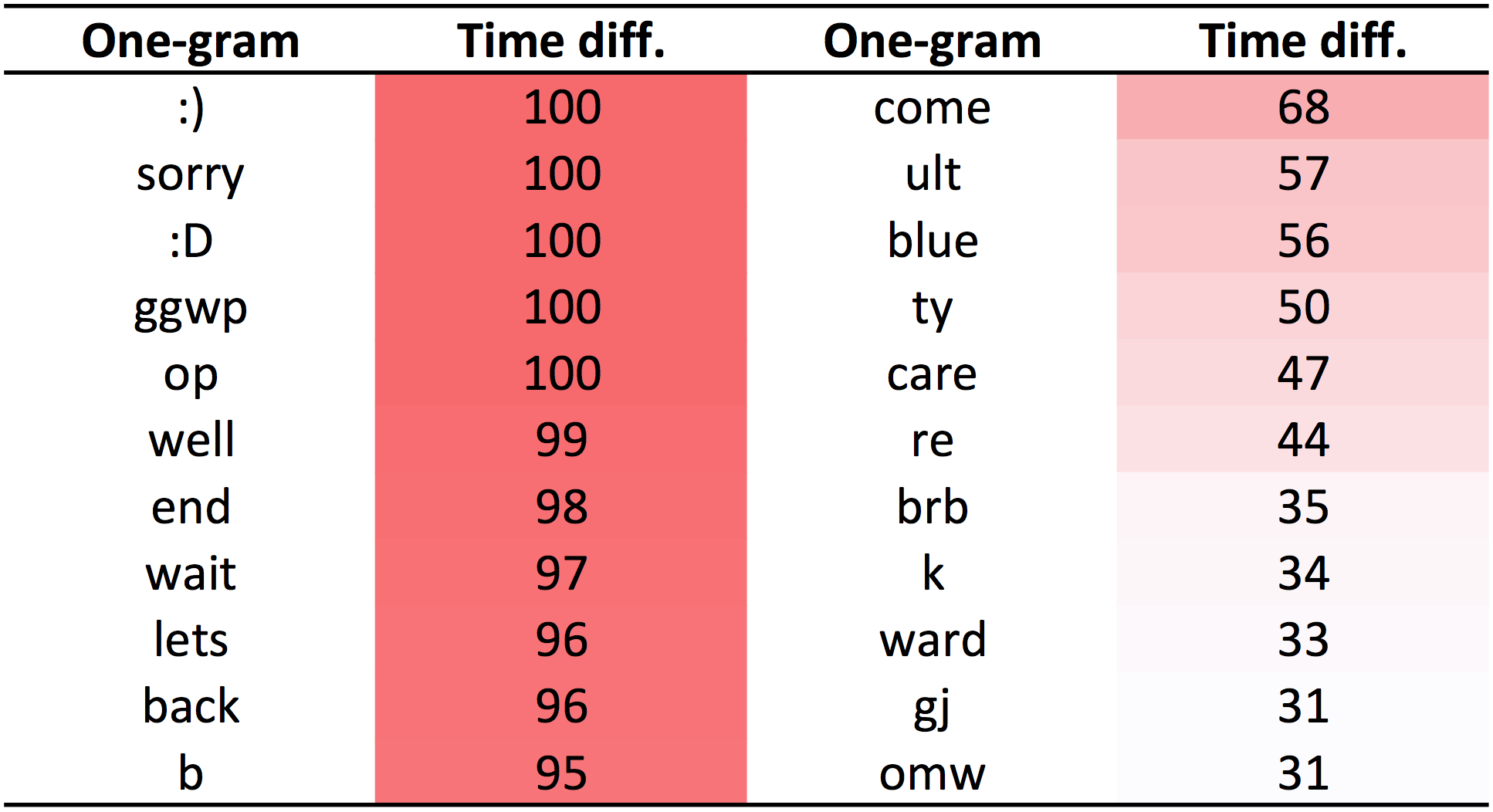}
  \caption{Time difference of last used time of uni-gram} 
  \vspace{-5mm} 
  \label{fig:timediff}
  \end{center}
\end{figure}

Figure~\ref{fig:timediff} lists the uni-grams with a time difference greater than 30.  I.e., words in the list are used later into the match by normal players.
Some interesting patterns are present in the results.

First, emoticons, particularly smile emoticons, are almost never used by toxic players.  
Second, apologies (e.g., `sorry') are also exclusively used by normal players. 
Third, some words for strategic team maneuvers (e.g., `come', `ult', `blue', `ward') are used by toxic players, but this ceases at some point during the match.  
Fourth, some words primarily used for communicating movements with partners in the same lane (e.g. `back', `b', `brb' (be right back), `omw' (on my way), `k' (okay)) are also used by toxic players, but again, after some point toxic players stop this form of communication.
Fifth, toxic players stop praising (e.g., `gj' (good job)) their teammates after some point in time.

All these findings reveal how toxicity is born during a match.  It seems to be a kind of phase transition.  They behave the same as normal players during the early stage of the match, but at some point they change their behavior.  After some point, they utter neither apologies nor praise to express their feelings, and also stop strategic communication with team members.  

By combining this finding with discriminative words of toxic players, we see the possibility for detecting a certain point that a player transitions to be toxic without using detailed in-game action logs, but just chat logs.  
Thus, linguistic analysis of toxic players shows not just how different they are and \emph{when} they become different as well.


\section{Conclusion and Future work}

In this work we have examined crowdsourced data from 590 thousand cases of accused toxic players in a popular match-based competition game, League of Legends.  We have performed a series of linguistic analyses to gain a deeper understanding of the role communication plays in the expression of toxic behavior.  
We have several interesting findings: a bi-modal distribution of chats during a match, a difference between temporal chat patterns between toxic and typical players, a list of discriminative uni- and bi-grams used by typical and toxic players as signatures of them, temporal patterns of the linguistic signature of toxic players, and a possible footprint of transitions from typical behavior to toxic behavior.  Our findings would be helpful to automatically detect and warn players who may turn to be toxic and thus save potential victims of toxic playing in advance.  

Finally, we suggest several directions for future work. 
First, is focusing on interaction between typical and toxic players.
In this work the unit of our analysis is a message, but we do not delve into the \emph{flow} of messages. 
Interaction analysis could reveal more clear narratives of how a player transitions to toxic behavior. 
Next, is building a pre-warning system to detect toxic playing earlier.
The main challenge here is to build a dictionary of words that are signs of toxic playing.
As we have seen a list of discriminative uni- and bi-grams of toxic and typical players, some bad words are also used by typical players as well. 
This behavior is prevalent in ``trash talk'' culture, and an important factor in immersing players in a competitive game~\cite{conmy2008trash}. 
Thus, any pre-warning system must be effective in detecting toxic playing while being flexible enough to allow for trash talk to avoid breaking the immersive gaming experience.
We believe that the signature of toxic and typical players we found is a first step for building the dictionary for a pre-warning system.

\bibliographystyle{abbrv}

\end{document}